%% file: ArxivRevised.tex
\documentclass[article,english,12pt]{article}
\usepackage[top=0.85in,left=1in,footskip=0.75in,marginparwidth=2in]{geometry}

\RequirePackage[OT1]{fontenc}
\RequirePackage{graphicx, color,amsmath, amssymb, amsthm, amsfonts, subfig}
\RequirePackage{mathpazo,natbib}
\RequirePackage{color,bm}
\RequirePackage{dsfont}
\RequirePackage{bigints}
\RequirePackage{enumerate}
\RequirePackage{bbm,lipsum}

\usepackage{tabularx}

\newcommand{\indic}{\mathbbm{1}}

\theoremstyle{plain}

\newtheorem*{theorem*}{Theorem}
\newtheorem{theorem}{Theorem}

\newtheorem*{definition*}{Definition}
\newtheorem{definition}{Definition}

\newtheorem*{prop*}{Proposition}

\newtheorem*{cor*}{Corollary}
\newtheorem{cor}{Corollary}

\renewcommand{\d}{\ensuremath{\mathrm{d}}}

\newcommand\numberthis{\addtocounter{equation}{1}\tag{\theequation}}

\newcommand\restr[2]{{
  \left.\kern-\nulldelimiterspace 
  #1 
  \vphantom{\big|} 
  \right|_{#2} 
  }}

\input{definitions.tex}

\begin{document}

\vspace*{0.35in}

\begin{flushleft}
{\Large
\textbf\newline{Integrability conditions for Compound Random Measures}
}
\newline
\\
Alan Riva Palacio and 
Fabrizio Leisen
\\
\bigskip
\bf{School of Mathematics, Statistics and Actuarial Sciences, University of Kent}\\
\bf{Sibson Building, Canterbury, Kent CT2 7FS}

\end{flushleft}

\section*{Abstract}Compound random measures (CoRM's) are a flexible and tractable framework for vectors of completely random measure. In this paper, we provide conditions to guarantee the existence of a CoRM.  Furthermore, we prove some interesting properties of CoRM's when exponential scores and regularly varying L\'evy intensities are considered.

\section{Introduction}\label{sec:intro}

Recently, a growing literature in Bayesian non-parametrics (BNP) proposed new priors which can take into account different features of the data, such as \textit{partial exchangeability}, see \cite{finetti}. In this case, one would like to consider different densities for different groups instead of a single common density for all the data. After the seminal paper of \cite{MacEachern},  the problem of modeling a finite number of dependent densities has become an active area of research in Bayesian non-parametrics. A common approach is to construct BNP priors based on functions of \textit{Completely Random Measures} (CRM's), see \cite{LP10}. For example, special attention has been given to the normalization of CRM's starting with the work of \cite{rlp}.  Roughly speaking, a CRM is a generalization of a subordinator, that is a process with independent increments and almost surely increasing paths; for a full account of CRM's see \cite{Kingman}. This property is very helpful to derive the Laplace functional transform which is the basis to derive some analytical quantities of interest such as posterior and predictive distributions or the Exchangeable Partition Probability Function (EPPF), see \cite{james}. To build more flexible priors in possibly higher dimensional spaces, vectors of dependent CRM's are constructed for example in \cite{LL}, \cite{LLS} and \cite{ZL} where respectively they build vectors of Poisson-Dirichlet and Dirichlet processes. These papers deal with the  L\'evy-Copula approach introduced in \cite{ContTank} to induce dependence among the components of the vector. In a similar fashion, \cite{NLP14a} introduce a vector of random probability measures where the dependence arises by virtue of a suitable construction of the Poisson random measures underlying the CRM's;
furthermore,  in the framework of survival analysis, \cite{NL14} introduce a new class of vectors of random hazard rate functions that are expressed as kernel mixtures of dependent CRM's. \cite{CameLP} focus on partial exchangeable models which arise from hierarchical specifications of CRM's. 

\textit{Compound random measures} (CoRM's), introduced by \cite{GL2017}, are a flexible and tractable framework for many dependent random measures including many of the superposition and L\'evy copula approaches.  They have recently been applied to modeling graphs for overlapping communities by \cite{caron16}.
\cite{GL2017} and \cite{GL2017b} described posterior sampling methods for a particular class of normalized compound random measure mixtures which exploits a representation of the Laplace transform of the CoRM through a univariate integral of a moment generating function. 

In this paper we aim to provide explicit existence conditions for CoRM's in order to guarantee the existence of the marginal L\'evy intensities. On the other hand, we prove that the resulting CoRM is well posed in the sense that it satisfies the usual integrability condition for multivariate L\'evy processes. Furthermore, this paper provides an interesting result for CoRM's when regularly varying L\'evy intensities are considered. The paper closes highlighting a representation on the multivariate L\'evy intensity of a CoRM when the score distribution is the result of marginal independent and identically distributed exponential scores. 

The outline of the paper is as follows. Section 2 will set the scene by introducing the basic definitions which are required in the CoRM setting. Section 3 is devoted to prove our main results. Section 4 deals with CoRM's built with regularly varying L\'evy intensities and exponential scores. Section 5 concludes.  

\section{Preliminaries}
Let $(\Omega, \mathcal{F}, \mathbb{P})$ be a probability space and $\mathbb{X}$ a Polish space with corresponding Borel $\sigma$-algebra $\mathcal{X}$. We denote by $\mathbb{M}_{\mathbb{X}}$ the space of boundedly finite measures on the measurable space $(\mathbb{X}, \mathcal{X})$ and by $\mathcal{M}_{\mathbb{X}}$ the associated Borel $\sigma$-algebra, see Appendix 2 in \cite{daley} for technical details. 
\begin{definition}\label{DefCRM}
A random measure $\mu$ on $\mathbb{X}$ is called a completely random measure (CRM) if for any $n > 1$ and disjoint sets $A_1, \dots, A_n \in  \mathcal{X}$ the random variables $\mu(A_1),\dots , \mu(A_n)$ are mutually independent.
\end{definition}
\noindent In the following we consider only CRM's without fixed jumps, namely CRM's of the form
$
\mu = \sum_{i=1}^\infty w_i\delta_{u_i}
$
for collections of random variables $\{w_i\}_{i=1}^\infty$ in $\re^+$ and $\{u_i\}_{i=1}^\infty$ in $\mathbb{X}$. Such CRM's can be characterized by their Laplace transform
\begin{align*}
\esp{e^{-\mu(f)}}=e^{-\int_{\re^+ \times \mathbb{X}} (1-e^{-sf(x)})\tilde{ \nu }(\d s ,\d x )} 
\end{align*}
where $\mu(f)=\int_\mathbb{X}f(x)\mu(\d x)$, $f:\mathbb{X}\rightarrow \re^+$ is such that $\mu(f)<\infty$ and $\tilde{\nu}(\d s , \d x)$ is a measure in $( \re^+\times \mathbb{X}$, $\mathcal{B}(\re^+)\otimes \mathcal{X} )$ such that
\begin{align}\label{Levycond}
\int_{\re^+\times X } \min\{1,s\} \tilde{\nu} (\d s,\d x) < \infty
\end{align}
for any bounded set $X\in \mathcal{X}$.
A measure $\tilde{\nu}$ satisfying the condition displayed in equation \eqref{Levycond} is called the \textit{L\'evy intensity} of $\mu$. We say that $\tilde{\nu}$ is homogeneous when $\tilde{\nu}(\d s, \d x) = \rho(\d s) \alpha(\d x)$ with $\rho$ a measure in $(\re^+, \mathcal{B}(\re^+))$ and $\alpha$ a measure in $(\mathbb{X},\mathcal{X})$.
The notion of a completely random measure can be generalized to higher dimensions in a similar fashion to Definition \ref{DefCRM}, see for instance \cite{GL2017}. As a result, we have a representation in terms of a Laplace functional transform. Precisely, for a vector of completely random measures $\pmb{\mu}=(\mu_1, \dots , \mu_d)$ on $\mathbb{X}$ we have that
\begin{align*}
\esp{e^{-\mu_1(f_1) - \dots - \mu_d(f_d) }}
=e^{-\int_{(\re^+)^d\times \mathbb{X}} \left(
1-e^{-s_1 f_1(x)-\dots - s_d f_d(x)}
\right)
\tilde{\nu}_d(\d \pmb{s},  \d x)}
\end{align*}
with $f_j:\mathbb{X}\rightarrow \re^+$, $j \in \{1,\dots, d\}$ such that $\mu_j(f_j) <\infty$, where for $g:\mathbb{X}\rightarrow \re^+$ we have $\mu_j(g)=\int_{\mathbb{X} } g(x) \mu_j(\d x)$.
The measure $\tilde{\nu}_d$ in $\left( (\re^+)^d \times \mathbb{X}, \mathcal{B}((\re^+)^d \otimes \mathcal{X} \right)$ must be such that
\begin{align}\label{Levycondmultiv}
\int_{(\re^+)^d \times \mathbb{X} } \min\{1,\|\pmb{s}\|\}\tilde{\nu}_d(\d \pmb{s},  \d x) < \infty
\end{align}
for any bounded set $X\in \mathcal{X}$;
we call $\tilde{\nu}_d$ a multivariate L\'evy intensity.
We set the notation
\begin{align*}
\nu_j(A,X) = \int_{(\re^+)^{d-1}}
\tilde{\nu}_d ( \d s_1,\dots , \d s_{j-1}, A, \d s_{j+1}, \dots, \d s_d, X)
\end{align*}
with $
j\in \{1, \dots , d\}$ and $A\in \mathcal{B}\left( \re^+ \right)$. We call $\nu_j$ the $j$-th marginal of the $d$-variate L\'evy intensity $\tilde{\nu}_d$; it follows that for each  $j\in \{1, \dots , \d\}$, $\mu_j$ has L\'evy intensity $\nu_j$. In this framework we can define the concept of \textit{Compound Random Measure} (CoRM). The following definition differs from the one in \cite{GL2017} since it takes into account the \textit{inhomogeneous} case, where the locations and associated weights in the CRM are not independent.
\begin{definition}
A \textit{Compound Random Measure} (CoRM) is a vector of CRM's whose L\'evy intensity is given by
\begin{align}\label{CoRM_def}
\tilde{\nu}_d(\d \pmb{s}, \d x) =  \int_{\re^+} z^{-d}h\left(
 \frac{ s_1}{z}, \dots , \frac{ s_d}{z}
\right)\d \pmb{s} \nu^\star (\d z, \d x)
\end{align}
where $h$, \textit{the score distribution}, is a d-variate probability density function and, $\nu^\star $, \textit{the directing L\'evy measure}, is a L\'evy intensity. 
\end{definition}
\noindent
By performing a simple change of variable we note that 
$$\int_{(\re^+)^d}z^{-d}h\left(
 \frac{ s_1}{z}, \dots , \frac{ s_d}{z}
\right)\d \pmb{s}=1.$$
Therefore,  $z^{-d}h\left(
 \frac{ s_1}{z}, \dots , \frac{ s_d}{z}
\right)$ can be seen as the density of a distribution function 
$$H\left(\frac{ \d s_1}{z}, \dots , \frac{\d s_d}{z}\right).$$ This allows to write the multivariate L\'evy intensity in equation \eqref{CoRM_def} as
\begin{align}\label{cormgen}
\tilde{\nu}_d(\d \pmb{s}, \d x) =  \int_{\re^+} H\left(
 \frac{ \d s_1}{z}, \dots , \frac{\d s_d}{z}
\right) \nu^\star (\d z, \d x).
\end{align}
To write the L\'evy intensity of a CoRM in terms of distribution functions rather than densities will be convenient for the results proved in the next section.

\section{Integrability conditions}
\noindent
The specification of a CoRM needs the initial choice of a score distribution and a directing L\'evy measure. Although this sounds straightforward, it is necessary to check that theses choices lead to a well defined CoRM.  Otherwise, the risk is to perform a Bayesian statistical analysis based on an ill-posed prior. In this section we look at two important aspects of Definition 2:
\begin{enumerate}
\item we provide conditions on the score distribution and the directing L\'evy measure for the existence of the marginal L\'evy intensities of a CoRM, see Theorem 1
and Corollary 1, 
\item we provide conditions on the score distribution and the directing L\'evy measure for the existence of the multivariate L\'evy intensity of a CoRM, see Theorem 2. 
\end{enumerate}
Summing up, Theorem 1 and Corollary 1 are focusing on the marginal existence of a CoRM. On the other hand,  Theorem 2 focuses on the global existence of a CoRM. The proofs of the theorems can be found in the supplementary material. 

\medskip

\begin{theorem}
Let $H$ be a $d$-variate score distribution and $\nu^\star$ a directing L\'evy measure defining a measure $\tilde{\nu}_d$ as in \eqref{cormgen} with corresponding marginals $\nu_j$ for $j\in \{1, \dots , \d\}$. Let $X$ be a bounded set in $\mathcal{X}$, then the measure $\nu_j$ satisfies the integrability condition (\ref{Levycond}) if and only if
\begin{align}\label{gencond1}
 \int_{ (0,1) \times X} \prob{ S_j  \geq \frac{1}{z}} \nu^\star (\d z, \d x) < \infty
\end{align}
and
\begin{align}\label{gencond2}
\int_{ [1,\infty)\times X} \prob{ S_j \leq \frac{1}{z}} z \nu^\star( \d z, \d x) < \infty.
\end{align}
Furthermore if the marginal score $H_j$ satisfies that
\begin{align}\label{particond1}
1-H_j\left( \frac{1}{z} \right) \leq z \;  \forall \; z\in (0,\epsilon) \text{ for some }  
\epsilon >0
\end{align}
and
\begin{align}\label{particond2}
\lim_{z \to \infty } z H_j\left(\frac{1}{z} \right)<\infty 
\end{align}
then conditions (\ref{gencond1}), (\ref{gencond2}) are satisfied with an arbitrary choice of the directing L\'evy measure $\nu^\star$. 
\end{theorem}

As set in Definition 2, we usually work with CoRM's given by a score with a probability density; in such case the following corollary to Theorem 1 follows.

\begin{cor}
If $S_j$ has probability density function $h_j$ then conditions (\ref{particond1}), (\ref{particond2}) reduce to 
\begin{align}\label{corcond1}
\lim_{z\to 0}\frac{h_j\left( \frac{1}{z} \right) }{z^2} < 1
\end{align}
and
\begin{align}\label{corcond2}
\lim_{\epsilon \to 0 }h_j \left( \epsilon \right)  < \infty.
\end{align}
\end{cor}
\noindent
The previous results concerned conditions for the marginals of a CoRM to be well defined, now we focus on such a result for the CoRM. 
For a score density function $h$ and directing L\'evy measure $\nu^\star$ to properly define a CoRM we need to check the condition (\ref{Levycondmultiv}) which takes the form
\begin{align}\label{Levycondcorm}
\int_{\re^+ \times X } \int_{ (\re^+)^d } 
\min \{ 1, \| \pmb{s} \| \}
h\left( \frac{s_1}{z}, \dots, 
\frac{s_d}{z} \right)
\frac{\d \pmb{s}}{z^d}
 \nu^\star( \mathrm{dz}, \d x) < \infty
\end{align}
for bounded set $X \in \mathcal{X}$.  As stated at the beginning of this section, in the next theorem we provide conditions on the score distribution and the directing L\'evy measure for the existence of the multivariate L\'evy intensity of a CoRM. This is equivalent to provide  conditions such that the latter inequality holds true. 
\begin{theorem}
Consider a CoRM which satisfies conditions (\ref{gencond1}) and (\ref{gencond2}) for each marginal $\nu_j$,  $j\in \{1, \dots , \d\}$, then the integrability condition (\ref{Levycondcorm}) is satisfied.
\end{theorem}
We conclude this section by providing two examples of the use of the previous results when considering Gamma and Beta distributed scored distributions.
\\
\textbf{Example 1: Gamma scores}
\\
We consider the marginal gamma score case. Let $h$ be the $d$-variate probability density of the score distribution; for $j \in \{1, \ldots , d\}$ we denote the $j$-th marginal density $h_j$ and let it correspond to a Gamma distribution with shape and rate parameters $\alpha_j,\; \beta_j$, i.e.
$$
h_j(s) = \frac{\beta_j^{\alpha_j} s^{\alpha_j-1}
e^{-\beta_j s}}{\Gamma(\alpha_j)}\indic_{\{s\in (0,\infty)\}}.
$$
We check the constraints (\ref{particond1}), (\ref{particond2}) by making use of Corollary 1 as we have probability densities. To check (\ref{corcond1}) we see that
$$
\lim_{s\to 0}\frac{h_j\left( \frac{1}{s} \right) }{s^2}
= \lim_{s\to 0}
\frac{\beta_j^{\alpha_j} 
e^{-\frac{\beta_j}{s}}}{\Gamma(\alpha_j)s^{\alpha_j+1}}=0
$$
and constraint (\ref{corcond2}) is satisfied for arbitrary L\'evy directing measure $\nu^\star$ whenever $\alpha_j \geq 1$, as in the examples presented in \cite{GL2017}. However for $\alpha_j < 1$ the associated CoRM will be well posed depending on the choice of $\nu^\star$. If for example we take the directing L\'evy measure to be the $\sigma$-stable, i.e.
\begin{align*}
\nu^\star (\d z, \d x) =\frac{\sigma}{z^{1+\sigma}}\d z \d x
\end{align*}
then constraint (\ref{gencond1}) in Theorem 1 can be reduced to
\begin{align*}
\int_0^1\int_{\frac{1}{z}}^\infty h_j(s)\frac{\sigma}{z^{1+\sigma}}\d s\d z 
=\int_1^\infty \int_{\frac{1}{s}}^1 h_j(s)\frac{\sigma}{z^{\sigma+1}}\d z \d s
=\int_1^\infty h_j(s)(s^\sigma -1)\d s<\infty,
\end{align*}
which is always satisfied since $h_j$ is a Gamma density. On the other hand, condition (\ref{gencond2}) in Theorem 1 becomes
\begin{align*}
\int_1^\infty \int_0^{\frac{1}{z}} h_j(s)\frac{\sigma z}{z^{1+\sigma}}\d s\d z 
= \int_0^1\int_1^{\frac{1}{s}} h_j(s)\frac{\sigma}{z^{\sigma}}\d z\d s
= \int_0^1 h_j(s)\sigma (s^{\sigma -1}-1)\d s <\infty
\end{align*}
which is not satisfied when $\alpha_j+\sigma<1$.
\\
\textbf{Example 2: Beta scores}
\\
In the setting as above, if the marginal scores are Beta distributed, i.e.
$$
h_j(s)=\frac{s^{\alpha_j-1}(1-s)^{\beta_j-1}}{B(\alpha_j,\beta_j)}\indic_{\{s\in (0,1)\}}
$$
then constraint (\ref{particond1}) becomes
$$
\lim_{s\to 0}\frac{h_j\left( \frac{1}{s} \right) }{s^2}
= \lim_{s\to 0}
\frac{(s-1)^{\beta_j-1}}{
s^{\alpha_j+\beta_j} B(\alpha_j,\beta_j)
}=0,
$$
so it is always satisfied; and condition (\ref{particond2}) is satisfied whenever $\alpha_j \geq 1$. We consider again a $\sigma$-stable L\'evy intensity for $\nu^\star$ when $\alpha_j<1$. Proceeding as in the previous example, constraint (\ref{gencond1}) becomes
\begin{align*}
\int_1^\infty \int_{\frac{1}{s}}^1 h_j(s)\frac{\sigma}{z^{\sigma+1}}\d z \d s=0<\infty
\end{align*}
so it always holds; and constraint (\ref{gencond2}) becomes
\begin{align*}
\int_0^1 h_j(s)\sigma (s^{\sigma -1}-1)\d s<\infty,
\end{align*}
which does not hold if $\alpha_j+\sigma <1$.

\section{Other interesting properties}

The aim of this section is to investigate two interesting properties of CoRM's. First, we focus on CoRM's which arise from regurlarly varying directing L\'evy measures. This result is motivated by the recent papers of \cite{CaronFox} and \cite{caron16}  which made use of regularly varying L\'evy measures to construct sparse random graphs. Second, we provide an explicit expression of the multivariate L\'evy intensity of a CoRM with independent exponential scores. This result is interesting when compared with Theorem 3.2 in \cite{ZL} and Corollary 2 in \cite{GL2017} which provide, respectively, the L\'evy copula representation and the Laplace exponent of CoRM's with independent exponential scores. The proofs of the theorems can be found in the supplementary material. \\

Both results deal with a $d$-variate CoRM given by an homogeneous directing L\'evy intensity $\nu^\star(\d z, \d x)=\rho^\star(\d z)\alpha(\d x)$. Therefore, the corresponding marginals can be written as $\nu_j(\d s, \d x)=\rho_j(\d s)\alpha(\d x)$ with $\rho_j$ a measure in $\left( \re^+, \mathcal{B}(\re^+) \right)$. 
\subsection{Regularly varying directing L\'evy measure}
In this section we focus on CoRM's given by a directing L\'evy measure that is regularly varying. We recall that a real valued function $L$ is slowly varying if $\lim_{t \to \infty}L(at)/L(t)=1\; \forall a>0$.
\begin{definition}
An homogeneous L\'evy measure $\rho^\star (\d z)\alpha(\d x)$ in $\re^+\times \mathbb{X}$ is said to be regularly varying if the tail integral $U^\star(y)=\int_y^\infty \rho^\star(\d s)$ is a regularly varying function, i.e. it satisfies
\begin{equation*}
U^\star(y) = L \left( \frac{1}{y} \right)\frac{1}{y^\sigma }
\end{equation*}
for some $\sigma \in [0,1)$ which we call the index and $L$ a slowly varying function.
\end{definition} 

The following Theorem highlights an interesting link between the directing L\'evy measure and the marginal L\'evy intensities in terms of the regularly varying property. 

\begin{theorem}
Consider a CoRM with an homogeneous directing L\'evy measure $\rho^\star(\d s)\alpha(\d x)$ such that the conditions of Theorem 2 are satisfied. If $\rho^\star$ is regularly varying with tail integral $U$ then the marginals $\rho_j$, $j=1,\dots,d$, are regularly varying.
\end{theorem}
\noindent \textbf{Example 3: $\sigma$-stable directing L\'evy measure}
\\
\noindent Consider a $\sigma$-stable directing L\'evy measure
$$
\nu^{\star}(\d s, \d x) = \frac{\sigma}{\Gamma(1-\sigma)s^{\sigma +1}}\d  s\d x.
$$
The related tail integral is
$$
U^\star(y)= \frac{1}{\Gamma(1-\sigma)y^\sigma}
$$
which is a regularly varying function with index $\sigma$ and slowly varying function $L(y)=\frac{1}{\Gamma(1-\sigma)}$. We see that the regularly varying tail integrals related to the CoRM construction arise as a factor of $U$, namely $U_j(y)=\esp{S_j^\sigma}U^\star(y)$ and from Theorem 3 the associated marginal tail integrals are regularly varying.

\subsection{Independent Exponential scores}

Consider a $d$-variate CoRM given by an homogeneous directing L\'evy measure $\nu^\star(\d z, \d x)=\rho^\star(\d z)\alpha(\d x) $ and a score distribution corresponding to $d$ independent standard exponential distributions, i.e. 
$$h(s_1,\ldots, s_d) = \prod_{i=1}^d e^{-s_i}.$$ 
We observe that each associated marginal takes the form 
$\nu_j(\d s, \d x)=f( s)\d s\alpha(\d x),$
where $f (s)=\int_0^\infty z^{-1}e^{-\frac{s}{z}} \rho^\star (\d z)$. The following Theorem provides a characterization for this class of CoRM's. 
\begin{theorem}
Consider a CoRM as described above; the corresponding $d$-variate L\'evy intensity 
$\tilde{\nu}_d(\d \pmb{s}, \d x)=\tilde{\rho}_d( \pmb{s})\d \pmb{s}\alpha(\d x)$ is such that
\begin{align*}
\tilde{\rho}_d(\pmb{s})=(-1)^{d-1}\restr{ \frac{\partial^{d-1}}{\partial s^{d-1}} f(s)}{s=s_1+\ldots +s_d}.
\end{align*}
\end{theorem}
\section{Conclusions}
In this paper, we proved some integrability condition for Compound Random Measures. The new findings can be useful to Statisticians which aim to use vectors of dependent completely random measures which arise from CoRM's with directing L\'evy measure and score distribution that have not been considered so far. Furthermore, in the homogeneous case, we proved that the marginal L\'evy intensities are regularly varying whenever a regularly varying directing L\'evy measure is considered. Finally, we provide a representation of the homogeneous CoRM's when the score distribution is the result of independent and identically distributed exponential distributions.   
\section*{Acknowledgments}
\noindent
Fabrizio Leisen was supported by the European Community's Seventh Framework Programme [FP7/2007-2013] under grant agreement no: 630677.
\section*{References}

\bibliographystyle{elsarticle-harv}

\section{Supplemetary Material: Proofs}

\noindent First, we recall the integrability condition for a L\'evy measure: 
\begin{align}\label{Levycond}
\int_{\re^+\times X } \min\{1,s\} \tilde{\nu} (\d s,\d x) < \infty
\end{align}
see Section 2 of the main document. 

Let $H$ and $\nu^\star$  be, respectively, a score distribution and a directing L\'evy measure which define a CoRM.  We denote with $H_j$, $j\in \{1, \dots , \d\}$, the $j$-th marginal of a $d$-dimensional score distribution $H$. A simple change of variable leads to the $j$-th marginal of a CoRM, namely
\begin{align*}
\nu_j \left( 
A, X
\right)
& = \int_{\re^+ \times X}\int_{A/z} H_j(
\d s )\nu^\star (\d z, \d x)
= \int_{\re^+}\int_{A \times X} \nu^\star
\left(
\frac{\mathrm{d}z}{s}, \d x \right)
H_j(\mathrm{d}s). \numberthis \label{use1}
\end{align*}
We can see the formula above as a mean. Let $S_j$ be a random variable with distribution $H_j$, then
\begin{align*}
\nu_j(A , X) = \esp{ \nu^\star \left( \frac{A}{S_j}, X \right) }
\numberthis \label{marg}
\end{align*}
for $A\in \mathcal{B}\left( \re^+ \right)$.
We use the last identity to give conditions for the marginal intensity $\nu_j$ to be a proper L\'evy intensity, i.e. a measure that satisfies the condition displayed in equation (\ref{Levycond}).
\subsection*{Proof of Theorem 1}

\noindent We recall the conditions stated in the first part of the theorem: 

\begin{align}\label{gencond1}
 \int_{ (0,1) \times X} \prob{ S_j  \geq \frac{1}{z}} \nu^\star (\d z, \d x) < \infty
\end{align}
and
\begin{align}\label{gencond2}
\int_{ [1,\infty)\times X} \prob{ S_j \leq \frac{1}{z}} z \nu^\star( \d z, \d x) < \infty,
\end{align}
and the conditions stated in the second part of the Theorem: 
\begin{align}\label{particond1}
1-H_j\left( \frac{1}{z} \right) \leq z \;  \forall \; z\in (0,\epsilon) \text{ for some }  
\epsilon >0
\end{align}
and
\begin{align}\label{particond2}
\lim_{z \to \infty } z H_j\left(\frac{1}{z} \right)<\infty.
\end{align}

\begin{proof}
We recall that $\nu^\star$ satisfies (\ref{Levycond}) since it is a L\'evy intensity. Using (\ref{marg}), condition (\ref{Levycond}) for $\nu_j$ becomes
\begin{align*}
& \esp{ \int_{\re^+ \times X}\min\{1, z\}\nu^\star\left(\frac{\d z}{S_j}, X\right)}
\\&
=
\esp{  \int_{ \left(0, \frac{1}{S_j} \right) \times X }  z\nu^\star (\d z, \d x) }
+
\esp{  \int_{ \left[\frac{1}{S_j},\infty \right)\times X } \nu^\star (\d z, \d x) }
< \infty. \numberthis \label{cond0}
\end{align*}
Therefore, $\nu_j$ satisfies (\ref{Levycond}) if and only if
\begin{align}\label{cond2}
& \esp{  \int_{ \left(0, \frac{1}{S_j} \right) \times X }  z\nu^\star (\d z, \d x) } < \infty
\end{align}
and
\begin{align}\label{cond1}
&\esp{  \int_{ \left[\frac{1}{S_j},\infty \right)\times X } \nu^\star (\d z, \d x) } < \infty.
\end{align}
The former can be decomposed using the Fubini-Tonelli theorem in
\begin{align*}
& \esp{  \int_{ \left( 0, \frac{1}{S_j} \right)  \times X }  z\nu^\star( \d z, \d x) } 
= \int_{\re^+ \times X} \prob{ S_j \leq \frac{1}{z}} z \nu^\star(\d z, \d x)
\\
& = \int_{ (0,1)\times X} \prob{ S_j \leq \frac{1}{z}} z \nu^\star(\d z, \d x)
+ \int_{ [1,\infty)\times X} \prob{ S_j \leq \frac{1}{z}} z \nu^\star( \d z, \d x).
\end{align*}
Condition \eqref{gencond2} ensures that the second term of the above equation is finite. It is easy to see that the first term is finite as well. Indeed, 
\begin{align*}
\int_{ (0,1)\times X} \prob{ S_j \leq \frac{1}{z}} z \nu^\star(\d z, \d x)
\leq 
\int_{ (0,1)\times X}  z \nu^\star(\d z, \d x)
 < \infty.
\end{align*}
On the other hand, the second term in \eqref{cond0} can be decomposed in
\begin{align*}
& \esp{  \int_{ \left[ \frac{1}{S_j},\infty \right) \times X }  \nu^\star (\d z, \d x)} 
= \int_{\re^+ \times X} \prob{ \frac{1}{z} \leq S_j } \nu^\star (\d z, \d x)
\\
& = \int_{ (0,1) \times X} \prob{ S_j  \geq \frac{1}{z}} \nu^\star (\d z, \d x)+ 
\int_{[1,\infty) \times X } \prob{ S_j  \geq \frac{1}{z}} \nu^\star (\d z, \d x).
\end{align*}
Condition \eqref{gencond1} ensures that the first term of the above equation is finite. It is easy to see that the second term is finite as well. Indeed,
\begin{align*}
\int_{[1,\infty) \times X} \prob{ S_j  \geq \frac{1}{z} } \nu^\star (\d z, \d x) & \leq \int_{ [1,\infty ) \times X}
\nu^\star (\d z, \d x) < \infty.
\end{align*}
Therefore, the first part of the theorem follows from (\ref{cond0}), (\ref{cond2}) and (\ref{cond1}).
\\
\noindent 
For the remaining part of the Theorem we use that \eqref{Levycond} is attained when considering the directing L\'evy measure $\nu^\star$. Indeed, if
\begin{align}
\lim_{z \to \infty } z\prob{ S_j \leq \frac{1}{z}}<\infty 
\end{align}
then as $\nu^\star$ is a L\'evy intensity
\begin{align}
\int_{ [1,\infty) \times X} \prob{ S_j \leq \frac{1}{z}} z 
\nu^\star(\d z, \d x)  < \infty.
\end{align}
so (\ref{cond2}) holds. And if there exists
$ \epsilon > 0 $  such that $1-H_j\left( \frac{1}{z} \right) \leq z \;  \forall \; z\in (0,\epsilon)  $
then 
\begin{align*}
\int_{ (0,1)\times X } \prob{ \frac{1}{z} \leq S_j } 
\nu^\star(\d z, \d x) 
<
\int_{ (0,1)\times X } z
\nu^\star(\d z, \d x)
< \infty,
\end{align*}
so (\ref{cond1}) also holds. From the first part of the theorem the CoRM marginal $\nu_j$ satisfies the integrability conditions for arbitrary $\nu^\star$.
\end{proof}

\subsection*{Proof of Corollary 1}

\noindent We recall the conditions stated in Corollary 1:

\begin{align}\label{corcond1}
\lim_{z\to 0}\frac{h_j\left( \frac{1}{z} \right) }{z^2} < 1
\end{align}
and
\begin{align}\label{corcond2}
\lim_{\epsilon \to 0 }h_j \left( \epsilon \right)  < \infty.
\end{align}

\begin{proof}
We define $f(z)=z-(1-H_j\left( \frac{1}{z} \right))$ and observe that $f(0^+)=0$ so the existence of $f'(0^+)>0$ implies (\ref{particond1}). As $S_j$ has a probability density we get that $f'(0^+)$ exists and (\ref{particond1}) is equivalent to $f'(0^+)>0$ which we write as
\begin{align*}
\lim_{z\to 0}\frac{h_j\left( \frac{1}{z} \right) }{z^2} < 1.
\end{align*}
Using the fundamental theorem of calculus we see that (\ref{particond2}) reduces to
\begin{align*}
\lim_{z \to \infty } z\prob{ S_j \leq \frac{1}{z}}
=\lim_{\epsilon \to 0 }h_j \left( \epsilon \right)  < \infty
\end{align*}
which is satisfied when $h_j$ is continuous at zero.
\end{proof}

\subsection{Proof of Theorem 2}

\begin{proof}

Denote 
$P_j = \{ \pmb{s}\in(\re^+)^d\, : \; \max\{s_1,\dots, s_d\}=s_j \}$ for  $j\in \{1, \dots , \d\}$; then, by using (\ref{use1}) and the fact that each $\nu_j$ is a L\'evy intensity we get that for any bounded set $X$ in $\mathcal{X}$
\begin{align*}
& \int_{\re^+ \times X } \int_{ (\re^+)^d } 
\min \{ 1, \| \pmb{s} \| \}
h\left( \frac{s_1}{z}, \dots, 
\frac{s_d}{z} \right)
\frac{\d \pmb{s}}{z^d}
 \nu^\star( \mathrm{dz}, \d x)
\\&=
\sum_{j=1}^d
\int_{\re^+ \times X } \int_{ P_j } 
\min \{ 1, \| \pmb{s} \| \}
h\left( \frac{s_1}{z}, \dots, 
\frac{s_d}{z} \right)
\frac{\d \pmb{s}}{z^d}
 \nu^\star( \mathrm{dz}, \d x) 
\\
& \leq 
\sum_{j=1}^d
\int_{\re^+ \times X } \int_{ P_j } 
\min \{ 1, \sqrt{d}s_j \}
h\left( \frac{s_1}{z}, \dots, 
\frac{s_d}{z} \right)
\frac{\d \pmb{s}}{z^d}
 \nu^\star( \mathrm{dz}, \d x) 
\\
& \leq 
\sum_{j=1}^d
\int_{\re^+ \times X } \int_{ (\re^+)^d} 
\min \{ 1, \sqrt{d}s_j \}
h\left( \frac{s_1}{z}, \dots, 
\frac{s_d}{z} \right)
\frac{\d \pmb{s}}{z^d}
 \nu^\star( \mathrm{dz}, \d x) 
\\&=
\sum_{j=1}^d
\int_{\re^+ \times X } \int_{ \re^+} 
\min \{ 1, \sqrt{d}s \}
h_j\left( \frac{s}{z}\right)
\frac{\d s}{z}
 \nu^\star( \mathrm{dz}, \d x) 
\\
&=
\sum_{j=1}^d
\int_{\re^+ \times X}
\min \{ 1, \sqrt{d}s \}
\nu_j(\d s, \d x)
\\&
\leq 
\sum_{j=1}^d
\int_{\re^+ \times X}
\min \{ \sqrt{d}, \sqrt{d}s \}
\nu_j(\d s, \d x)
\\&=
\sqrt{d}
\sum_{j=1}^d
\int_{\re^+ \times X}
\min \{ 1, s \}
\nu_j(\d s, \d x)
 < \infty.
\end{align*}
\end{proof}

\subsection*{Proof of Theorem 3} 
We recall that for the case at hand

\begin{equation}\label{def3ref}
U^\star(y) = L \left( \frac{1}{y} \right)\frac{1}{y^\sigma }
\end{equation}
is a tail integral.
\begin{proof}
We note that equation (\ref{marg}) implies that
\begin{align*}
\rho_j(A)=\esp{\rho^\star\left(\frac{A}{S_j}\right) }
\end{align*}
It follows that the marginals of the CoRM are given by
\begin{align*}
U_j(y) & =  \rho_j \left( (y,\infty) \right) 
\\&= \esp{ U^\star\left(
\frac{y}{S_j}
\right)}
\\&
=\esp{
L\left(
\frac{S_j}{y}
\right) \left(
\frac{S_j}{y} \right)^\sigma
}
\\&=
\esp{L\left(
\frac{S_j}{y}
\right)
S_j^\sigma
}
\frac{1}{y^\sigma	}.
\end{align*}
Hence, it is enough to check if the function $l(z)= \esp{ L(S_j z)S_j^\sigma }$ is slowly varying for $L$ a slowly varying function. Let $a>0$, we need to check
\begin{align*}
\lim_{t\to \infty} \frac{ l(at) }{l(t)}
& = 
\lim_{t\to \infty} 
 \frac{ \esp{ L( a t S_j )S_j^\sigma } }{\esp{ L( t S_j )S_j^\sigma }} =1.
\end{align*}
For a fixed $\epsilon > 0 $ we can choose $t_0$ such that $\forall u > t_0$
\begin{align*}
\left| L(a u)/L( u) - 1 \right| < \frac{\epsilon}{2}, 
\end{align*}
since $L$ is slowly varying.  Then for $t>t_0$
\begin{align*}
&\left|
\frac{ \esp{ L( a t S_j )S_j^\sigma } }{\esp{ L( t S_j )S_j^\sigma }} -1
\right|
=
\left|
\frac{ \esp{ S_j^\sigma \left( L( a t S_j )- L( t S_j ) \right) }
 }{\esp{ L( t S_j )S_j^\sigma }} 
\right|
\\
&\leq
\frac{ \esp{ \indic_{ \left\{ S_j > \frac{t_0}{t} \right\}} S_j^\sigma \left| L( a t S_j )- L( t S_j ) \right| }
 }{\esp{ L( t S_j )S_j^\sigma }} 
+
\frac{ \esp{ \indic_{ \left\{ S_j \leq \frac{t_0}{t} \right\}} S_j^\sigma \left| L( a t S_j )- L( t S_j ) \right| }
 }{\esp{ L( t S_j )S_j^\sigma }} 
\\
&< 
\frac{ \esp{ \indic_{ \left\{ S_j > \frac{t_0}{t} \right\}} S_j^\sigma \frac{\epsilon}{2} L(t S_j)}
 }{\esp{ L( t S_j )S_j^\sigma }} 
+
\frac{ \esp{ \indic_{ \left\{ S_j \leq \frac{t_0}{t} \right\}} S_j^\sigma \left| L( a t S_j )- L( t S_j ) \right| }
 }{\esp{ L( t S_j )S_j^\sigma }} 
\\
&
< \frac{\epsilon}{2} + 
\frac{ \esp{ \indic_{ \left\{ S_j \leq \frac{t_0}{t} \right\}} S_j^\sigma \left| L( a t S_j )- L( t S_j ) \right| }
 }{\esp{ L( t S_j )S_j^\sigma }}  
\\&=
\frac{\epsilon}{2} + 
\frac{ \esp{ \indic_{ \left\{ S_j \leq \frac{t_0}{t} \right\}} S_j^\sigma t^\sigma \left| L( a t S_j )- L( t S_j ) \right| }
 }{U_j(\frac{1}{t})}  
\\
&=
\frac{\epsilon}{2} + 
\frac{ \int_{(0, \frac{t_0}{t}]} s^\sigma t^\sigma \left| L( a t s )- L( t s ) \right| H_j(\d s) 
}{U_j(\frac{1}{t})}  
\\
&=
\frac{\epsilon}{2} + 
\frac{ \int_{(0, t_0]} u^\sigma \left| L( a u )- L( u ) \right| H_j \left( \frac{ \d u}{t} \right) 
}{U_j(\frac{1}{t})}   \numberthis \label{cuentateo}
\end{align*}
We observe that 
$
\lim_{x\to 0} U_j(x)= \infty.
$ since $U_j$ is a tail integral. From \eqref{def3ref} it follows that
$
\lim_{ x \to 0} x^\sigma \left( 
L(ax) -L(x)
\right)=0
$. Hence, the function $g(x)= x^\sigma \left( 
L(ax) -L(x)
\right)$ is bounded in $[0,t_0]$ by a constant $K_{1,t_0}$. Finally we observe  that for $t>t_0$
$$
\int_{(0,t_0]} H_j \left( \frac{ \d u}{t} \right)
<
\int_{(0, 1]} H_j \left( \d u \right)
\leq 1.
$$
We set $t_1>t_0$ such that for $u>t_1$
$$
\frac{ 2 K_{1,t_0 }}{\epsilon}  < U_j(1/u).
$$
Choosing $t>t_1$ we get
\begin{align*}
\frac{ \int_{(0, t_0]} u^\sigma \left| L( a u )- L( u ) \right| H_j \left( \frac{ \d u}{t} \right) 
}{U_j(\frac{1}{t})}< \frac{K_{1,t_0} }{U_j(\frac{1}{t})}<\frac{\epsilon}{2} 
\end{align*}
It follows from \eqref{cuentateo} that
$$
\left|
\frac{ \esp{ L( a t S_j )S_j^\sigma } }{\esp{ L( t S_j )S_j^\sigma }} -1
\right| < \epsilon.
$$
Consequently, $l$ defined above is slowly varying, implying that the marginal tail integral $U_j$ is regularly varying.
\end{proof}

\subsection{Proof of Theorem 4}

\begin{proof}
It is straightforward to see that
\begin{align*}
\tilde{\rho}_d(\pmb{s})= \int_0^\infty z^{-d}h\left(
\frac{s_1}{z}, \cdots , \frac{s_d}{z}
\right)\rho^\star(\d z)
= \int_0^\infty z^{-d}
e^{-\frac{s_1+\ldots + s_d}{z}}
\rho^\star(\d z).
\end{align*}
From Example 1 in the main document, we know that, for arbitrary $\rho^\star$ and $d\in \mathbb{N}\setminus \{0\}$, the previous integral is finite. Therefore for $s \neq 0$
\begin{align*}
\int_0^\infty
\left|
\frac{\partial^j}{\partial s^j}
z^{-1}e^{-\frac{s}{z}}
\right|
\rho^\star (\d z)
 &=
 \tilde{\rho}_{j+1} (s,0,\ldots,0) < \infty
\end{align*}
and this concludes the proof since using the Dominated Convergence Theorem we can take the derivative under the integral sign as follows
\begin{align*}
(-1)^{d-1}\restr{ \frac{\partial^{d-1}}{\partial s^{d-1}} f(s)}{s=s_1+\ldots +s_d}
\;
&= \;  
(-1)^{d-1}\restr{ 
\int_0^\infty 
\frac{\partial^{d-1}}{\partial s^{d-1}} \left(
z^{-1}e^{-\frac{s}{z}}
\right)
\rho^\star ( \d z)
}{s=s_1+\ldots +s_d}
\\&
= \; \int_0^\infty z^{-d}
e^{-\frac{s_1+\ldots + s_d}{z}}
\rho^\star(\d z).
\end{align*}
\end{proof}
\end{document}

%% file: definitions.tex
\newcommand{\bb}[1]{\mathbb{#1}}

\newcommand{\sip}{\bb{P}}

\newcommand{\se}{\ensuremath{\bb{E}}}

\newcommand{\re}{\ensuremath{\mathbb{R}}}

\newcommand{\prob}[1]{\ensuremath{\sip\! \left[ #1 \right]}}

\newcommand{\esp}[1]{\ensuremath{\se\! \left[ #1 \right]}}

